\documentclass[preprint,epsfig,eqsecnum,aps]{revtex4}
\usepackage{graphicx}
\usepackage{epsfig}
\begin{document}
%\draft
\title{New results for the fully renormalized proton-neutron  quasiparticle random phase approximation}

\author{C. M. Raduta$^{b)}$, A. A. Raduta$^{a), b)}$}

\address{
$^{a)}$Department of Theoretical Physics and Mathematics,Bucharest University, POBox MG11,
Romania}
\address{$^{b)}$Institute of Physics and Nuclear Engineering, Bucharest, POBox MG6, Romania}

\begin{abstract}
A many-body Hamiltonian describing a system of Z protons and N neutrons moving in
spherical shell model mean field and interacting among themselves through  proton-proton and
neutron-neutron pairing and a dipole-dipole proton-neutron interaction of both particle-hole and
particle-particle type, is treated within a fully renormalized (FR) pnQRPA approach.
Two decoupling schemes are formulated. One of them decouples the equations of
motion of
particle total number conserving and non-conserving operators. One ends up with
two very simple
dispersion equations for phonon operators which are formally of Tamm-Dancoff types.
For excitations in the (N-1,Z+1) system, Ikeda sum rule is fully satisfied
provided
the BCS equations are renormalized as well and therefore solved at a time with
the FRpnQRPA equations. Next, one constructs two operators ${\cal R}^{\dagger}_{1\mu}$,
${\cal R}_{1,-\mu}(-)^{1-\mu}$ which commutes with the particle total number
conserving operators, ${\cal A}^{\dagger}_{1\mu}$ and
${\cal A}_{1,-\mu}(-)^{1-\mu}$, and moreover could be renormalized so that
they become bosons. Then, a phonon operator is built up as a linear
combination of these four operators.
The FRpnQRPA equations are written down for this complex phonon operator and
the ISR is calculated analytically. This formalism allows for an unified
description of the dipole excitations
in four neighboring nuclei (N-1,Z+1),(N+1,Z-1),(N-1,Z-1),(N+1,Z+1). The phonon vacuum describes the
(N,Z) system ground state.
\end{abstract}
%{\tt$\backslash$\string pacs\{\}} should always be input,
%even if empty.}

\maketitle

\section{Introduction}
\label{sec:level1}
Understanding nuclear physics phenomena in terms of single particle motion in 
a mean field and nucleon-nucleon interaction
is one of the most appealing aims of theoretical nuclear physics.
Along the latest few decades many progresses have been obtained in the many body treatment of nuclear system.
 Among the chief achievements one distinguishes the Hartree-Fock,
BCS and random phase approximation (RPA)\cite{GreiMar,Ring}. The last mentioned approach can be
formulated on the top of either a HF or a BCS ground state. The second version is
conventionally called quasiparticle random phase approximation (QRPA).
Among many successes of many body theories are the description of the ground state
properties, of the excited states, including those of high spin of ground as well as of
excited bands, of the electromagnetic transitions, and of the reaction mechanisms.

The proton neutron interaction has been also treated. For example the proton-neutron
pairing interaction is currently investigated for nuclei close to the drip line.
It is still an open question whether there exists a T=0 proton-neutron pairing
\cite{Machia} phase,
although there are some claims that the strong back-bending seen for $^{52}$Fe
could be explained by breaking a proton-neutron T=0 pair \cite{Pov}.

The dipole proton-neutron interaction in the particle-hole ($ph$) as well as the
particle-particle ($pp$) channels have been treated within the pnQRPA formalism
in order to get a quantitative description of the double beta decay with two neutrinos in the
final state, $2\nu\beta\beta$\cite{Su,Fa}. Of course, this phenomenon is very interesting by its own, but
 exhibits a special attraction for theoreticians  having in view that it provides a test for the  nuclear matrix elements which 
are also used in the neutrinoless double beta decay ($0\nu\beta\beta$)
calculations.
Indeed, the discovery of $0\nu\beta\beta$ process would answer a fundamental question whether neutrino
is a Majorana or a Dirac particle.

Standard pnQRPA calculations based on $ph$ two body interaction yields a too large rate
for the $2\nu\beta\beta$ process. In Ref.\cite{Cha} it was pointed out that the
$\beta^+$ transition  matrix element is very sensitive to the $pp$ interaction strength.
Since the double beta decay has a branch which could be looked at as the
hermitian conjugate
matrix element of the $\beta^+$ virtual transition of the daughter nucleus to
an dipole $1^+$ state in the intermediate odd-odd nucleus, many groups working on
$2\nu\beta\beta$ decay included the $pp$ two body dipole interaction
in the pnQRPA calculations. By contrast to the $ph$
interaction, which is repulsive, the $pp$ interaction has an attractive character.
 Since such an interaction is not considered in the
mean field equations, the approach fails at a critical value of the interaction strength,
$g_{pp}$. Before this value is reached the Gamow-Teller transition amplitude, denoted
by $M_{GT}$, is rapidly decreasing and after a short interval is becoming equal to zero.
The experimental data for the amplitude $M_{GT}$ is met for a value of $g_{pp}$ close to that for which
$M_{GT}$ vanishes and moreover close to the critical value of $g_{pp}$ where pnQRPA breaks
down.
It is obvious that in this region of $g_{pp}$, the pnQRPA results are not
stable
against adding interaction terms which are not encountered by the adopted many body approach.
In order to restore the ground state stability, one needs to make  a higher RPA calculation.
The first formalism devoted to this feature includes anharmonicities through
the boson expansion technique
\cite{Rad1,Rad2}. Another method is the renormalized pnQRPA procedure (pnRQRPA)
\cite{Su1}
which keeps the harmonic picture but renormalizes the bosons by effects coming from some
 terms of the commutator algebra which are not taken into account in the standard pnQRPA.

In a previous paper\cite{Rad3} we have shown that the pnRQRPA does not include the higher RPA effects in a
consistent way. Indeed, in Ref.\cite{Su1} only the two quasiparticle dipole operators are renormalized
due to the non-vanishing average of the quasiparticle number operators in the renormalized
QRPA ground state.  In Ref.\cite{Rad3} we showed that having non-vanishing
proton and neutron quasiparticle numbers, the scattering terms are also renormalized so that they finally satisfy bosonic
commutation relations. Thus, a new phonon operator could be defined which includes, in addition,
the scattering terms. We baptized the new renormalization formalism as a fully renormalized
proton-neutron quasiparticle random phase approximation (FRpnQRPA).
The FRpnQRPA equations determining the phonon operator amplitudes have twice as
much solutions, $2N_s$, as
the standard pnRQRPA equation. For half of these solutions one expects that the amplitude
of the scattering terms prevails. In the quoted reference we pointed out which 
contribution is coming from the $ph$ and which comes from the $pp$ interactions in building up
 the new modes. It was shown that for charge conserving excitations these
 modes are spurious\cite{Bla}.
 In a later publication we treated the scattering terms semi-classically and
 the harmonic mode defined there describes the wobbling motion of the isospin
 degrees of freedom\cite{Rad4}.

 It is worth mentioning that both renormalization approaches, pnRQRPA and FRpnQRPA,
 violates the   Ikeda sum rule (ISR)\cite{Ike}, saying that the difference of the
 total $\beta^-$ and $\beta^+$ strengths for a single beta minus emitter is equal to
 3(N-Z).
 Aiming at conciliating  the two features, renormalization and having ISR
 satisfied, recently \cite{Ro} a new phonon operator was  defined which includes also
 the scattering terms but commutes with the particle total number operator.
 The resulting formalism is called  fully renormalized pnQRPA,i.e. by the name
  that we adopted  four years before.
A distinct way to renormalize the RPA equations
without modifying the number of solutions and moreover without deriving an explicit expression for the phonon operator was formulated in Ref.\cite{Schu}. The method is known under the name SCRPA (self consistent RPA).

The present paper continues the investigation we started in Ref\cite{Rad3},
addressing the following issues.
We want to see to what extent it is possible, in a quasiparticle BCS framework,
to separate a harmonic boson operator conserving the particle total number from
the components which do not commute with the total number operator.
We shall see that the phonon operator conserving the particle total number,
 treated in a restricted $N_s$ dimensional space, can be embedded into
an operator space acting on a $2N_s$ dimensional space and moreover has
exactly the same properties as
the phonon operator we have previously introduced in Ref.\cite{Rad3}.
Also, we aim at treating on an equal footing the  particle-hole and deuteron like dipole excitations.
For both cases the FRpnQRPA equations are written down explicitly. Analytical expressions for
the renormalization factors will be derived. 
An important question addressed in this paper is whether
having a  particle total number projected phonon operator assures
that the ISR is fulfilled.
Another problem treated in the present paper is the following one.
In the FRpnQRPA approach, the phonon operator is a linear combination of
the renormalized operators (see the notations from the next sections)
$A^{\dagger}_{1\mu}(pn), A_{1,-\mu}(pn)(-)^{1-\mu}, B^{\dagger}_{1\mu}(pn),
 B_{1,-\mu}(pn)(-)^{1-\mu}$, neglecting the commutators of $A$ and $B$ operators.
 Here we define two renormalizable operators ${\cal R}^{\dagger}_{1\mu}(pn),{\cal R}_{1,-\mu}(pn)
 (-)^{1-\mu}$ which are exactly commuting with the operators ${\cal A}^{\dagger}_{1\mu}(pn)$ and
$ {\cal A}_{1,-\mu}(pn)(-)^{1-\mu}$, where the latter operators are preserving the total number of particles.
 The phonon operator is defined as a linear combination of
 the ${\cal A}$ and ${\cal R}$ operators and explicit equations for the phonon amplitudes and
 energies will be derived.

The objectives sketched above will be accomplished according to the following plan.
In Section 2 we briefly review the main ingredients of the FRpnQRPA formulated in
Ref.\cite{Rad3}. In Section 3 a consistent decoupling scheme is defined which results in providing
two independent sets of Tamm-Dancoff equations describing excitations of the (N,Z) system
in the nuclei (N-1,Z+1) and (N+1,Z+1), respectively.
For the first type of excitation the Ikeda sum rule is fulfilled.
In Section 4, a complex phonon operator is defined, describing in a unified fashion
states in  the systems  (N-1,Z+1), (N+1,Z-1), (N+1,Z+1), (N-1,Z-1), respectively.
Finally, in Section 5 the important results are summarized and some conclusions are drawn.

\section{Brief review of the fully renormalized pnQRPA formalism}
\label{sec:level2}
The single and double beta Gamow-Teller  transitions can be described by the following
many body model Hamiltonian:

\begin{eqnarray}
H&=&\sum_{\tau;jm}(\epsilon_{\tau jm}-\lambda_{\tau}c^{\dagger}_{\tau jm}
c_{\tau jm})-\sum_{\tau;j,j'}\frac{G_{\tau}}{4}P^{\dagger}_{\tau j}P_{\tau j'}
\nonumber\\&+&
2\chi \sum_{pn,p'n';\mu} \beta^-_{\mu}(pn)\beta^+_{-\mu}(p'n')(-)^{\mu}
-2\chi_1\sum_{pn,p'n';\mu} P^-_{1\mu}(pn)P^+_{1,-\mu}(p'n')(-)^{\mu}.
\end{eqnarray}
where $c^{\dagger}_{\tau jm}$ ($c_{\tau jm}$) denotes the creation (annihilation) operator for
a nucleon of type $\tau$ (=p,n) in a spherical shell model $|nljm\rangle$.
The time reversed state corresponding to $|nljm\rangle$ is
$\widetilde{|nljm\rangle}=|nlj-m\rangle(-)^{j-m}$.
For the sake of
simplifying the notation  here only
the quantum numbers $j,m$ are specified.
Also, the following notations have been used:
\begin{eqnarray}
\beta^-_{\mu}(pn)&=&\sum\langle pjm|\sigma_{\mu}|nj'm'\rangle c^{\dagger}_{pjm}c_{nj'm'}=
-\frac{\hat {j}_p}{\hat {1}}\langle pj||\sigma||nj'\rangle
\left[c^{\dagger}_{pj}c_{\widetilde{nj'}}\right]_{1\mu}, \nonumber\\
P^-_{1\mu}(pn)&=&\sum\langle pjm|\sigma_{\mu}|nj'm'\rangle c^{\dagger}_{pjm}
c^{\dagger}_{\widetilde{nj'm'}}=
\frac{\hat {j_p}}{\hat {1}}\langle pj||\sigma||nj'\rangle\left[c^{\dagger}_{pj}
c^{\dagger}_{\widetilde{nj'}}\right]_{1\mu},
\nonumber\\
\beta^+_{\mu}(pn)&=& \left(\beta^-_{-\mu}(pn)\right)^{\dagger} (-)^{\mu},\;
\;\;\;\;P^+_{1\mu}(pn)=\left(P^-_{1,-\mu}(pn) \right)^{\dagger} (-)^{\mu}.
\end{eqnarray}
\vskip0.5cm

In what follows, sometimes the notations are simplified even more and the set of quantum numbers for a
single $\tau$ particle state will be denoted by $\tau$,
the other quantum numbers being mentioned only if necessary.

The Hamiltonian H is treated first by the BCS formalism which defines the quasiparticle
representation through the Bogoliubov-Valatin (BV) transformation:
\begin{eqnarray}
a^{\dagger}_{\tau jm}&=&U_{j_{\tau}}c^{\dagger}_{\tau jm} -s_{jm}V_{j_{\tau}}c_{\tau j-m}
\nonumber\\
           U_{j_{\tau}}^2+V_{j_{\tau}}^2&=&1,\;\; s_{jm}=(-)^{j-m},\;\;\tau=p,n
\end{eqnarray}
The images of the operators $\beta^{\pm}_{\mu}$ and $P^{\pm}_{1\mu}$
due to the BV transformations are:
\begin{eqnarray}
\beta^-_{\mu}(k)&=&\xi_k A^{\dagger}_{1\mu}(k) +\bar{\xi}_kA_{1,-\mu}(k)
(-)^{1-\mu}+\eta_kB^{\dagger}_{1mu}-\bar{\eta}_kB_{1,-\mu}(k)(-)^{1-\mu},
\nonumber\\
\beta^+_{\mu}(k)&=&-\left[\bar{\xi}_k A^{\dagger}_{1\mu}(k)+\xi_k A_{1,-\mu}(k)
(-)^{1-\mu}-\bar{\eta}_k B^{\dagger}_{1mu}+\eta_k B_{1,-\mu}(k)(-)^{1-\mu}\right],
\nonumber\\
P^-_{\mu}(k)&=&\eta_kA^{\dagger}_{1\mu}(k) -\bar{\eta}_kA_{1,-\mu}(k)
(-)^{1-\mu}-\xi_kB^{\dagger}_{1mu}-\bar{\xi}_kB_{1,-\mu}(k)(-)^{1-\mu},
\nonumber\\
P^+_{\mu}(k)&=&\left[-\bar{\eta}_kA^{\dagger}_{1\mu}(k) +\eta_kA_{1,-\mu}(k)
(-)^{1-\mu}-\bar{\xi}_kB^{\dagger}_{1mu}-\xi_kB_{1,-\mu}(k)(-)^{1-\mu}\right].
\end{eqnarray}
where the operators $A^{\dagger}$,$A$,$B^{\dagger}$,$B$,  are the  two quasiparticle and 
quasiparticle density dipole operators, respectively:
\begin{eqnarray}
A^{\dag}_{1\mu}(pn)&=&\sum C^{j_p\;j_n\;1}_{m_p\;m_n\;\mu}a^{\dag}_{pj_pm_p}
a^{\dag}_{nj_nm_n},\;\; A_{1\mu}(pn)=\left(A^{\dag}_{1\mu}(pn) \right)^{\dag},
\nonumber\\
B^{\dag}_{1\mu}(pn)&=&\sum C^{j_p\;j_n\;1}_{m_p\;-m_n\;\mu}a^{\dag}_{pj_pm_p}
a_{nj_nm_n}(-)^{j_n-m_n},\;\; B_{1\mu}(pn)=\left(B^{\dag}_{1\mu}(pn) \right)^{\dag}.
\end{eqnarray}
The factors $\xi, \bar{\xi},\eta \bar{\eta}$ have the expressions:

\begin{eqnarray}
{\xi}_k &=& \frac{{\hat j}_p}{{\hat 1}}\langle j_p||\sigma ||j_n\rangle
U_{j_p}V_{j_n},\;\;\;\;
\bar{\xi}_k=\frac{{\hat j}_p}{\hat{1}}\langle j_p||\sigma ||j_n\rangle
V_{j_p}U_{j_n},\;\;\hat{j}=\sqrt{2j+1},\nonumber\\
\eta_k &=& \frac{{\hat j}_p}{\hat{1}}\langle j_p||\sigma ||j_n\rangle
U_{j_p}U_{j_n},\;\;\;\;
\bar{\eta}_k=\frac{{\hat j}_p}{\hat{1}}\langle j_p||\sigma ||j_n\rangle
V_{j_p}V_{j_n}.
\end{eqnarray}
It is worth mentioning that throughout this paper the Rose convention for the reduced matrix element is adopted
\cite{Rose}.
Within the proton-neutron quasiparticle random phase approximation (pnQRPA),
one assumes that the two quasiparticle dipole operators $A^{\dagger}_{1\mu},A_{1\mu'}$
satisfy quasi-bosonic commutation relations, while the quasiparticle  density dipole operator
commute with each others. Also the $A$ operators commute with any of $B^{\dagger}$ and $B$
operators.
The renormalized pnQRPA approximate the commutator
$\left[A_{1\mu},A^{\dagger}_{1\mu}\right]$ by keeping from its exact expression
only the $C$ number term and the monopole term. Moreover, replacing the
adopted expression for the commutator
by its average on the pnQRPA ground state, unknown yet, the operators $A$ and
$A^{\dagger}$ can be renormalized so that the new operators obey bosonic
commutation relation. Note that other commutator equations, are kept as in
the standard pnQRPA, i.e. they are taken equal to zero.

The idea advanced in Refs. \cite{Rad3} was that also the
proton-neutron quasiparticle density dipole operators can be renormalized 
so that finally they
are gifted with boson properties.
Therefore, the mutual commutation relations assumed in Ref.\cite{Rad3} are:
\begin{eqnarray}
\left[A_{1\mu}(k),A^{\dagger}_{1\mu'}(k')\right] &\approx &\delta_{k,k'}\delta{\mu,\mu'}
\left[1-\frac{{\hat N}_n}{{\hat j}^2_n}-\frac{{\hat N}_p}{{\hat j}^2_p}\right],
\nonumber\\
 \left[B^{\dagger}_{1\mu}(k),A^{\dagger}_{1\mu'}(k')\right] &\approx &
\left[B^{\dagger}_{1\mu}(k),A_{1\mu'}(k')\right] \approx 0,
\nonumber\\
 \left[B_{1\mu}(k),B^{\dagger}_{1\mu'}(k')\right] &\approx &\delta_{k,k'}\delta_{\mu,mu'}
\left[\frac{{\hat N}_n}{{\hat j}^2_n}-\frac{{\hat N}_p}{{\hat j}^2_p}\right]
,\;k=(j_p,j_n).
\label{comm2}
\end{eqnarray}
Let us denote by $C^{(1)}_{j_p,j_n}$ and  $C^{(2)}_{j_p,j_n}$ the averages
of the r.h. s. of the first and third commutation relations (\ref{comm2}) on the pnQRPA
vacuum state ($|0\rangle$), respectively.  Then, the renormalized operators are:
\begin{eqnarray}
\bar{A}^{\dagger}_{1\mu}(k) &=& \frac{1}{\sqrt{C^{(1)}_k}}A^{\dagger}_{1\mu}(k),\;
\bar{A}_{1\mu}(k)=\frac{1}{\sqrt{C^{(1)}_k}}A_{1\mu}(k),\nonumber\\
\bar{B}^{\dagger}_{1\mu}(k) &=& \frac{1}{\sqrt{|C^{(2)}_k|}}B^{\dagger}_{1\mu}(k),\;
\bar{B}_{1\mu}(k)=\frac{1}{\sqrt{|C^{(2)}_k|}}B_{1\mu}(k).
\end{eqnarray}
Recalling the specific RPA convention to  replace the operator commutators by their
average on the ground state, one readily obtains that the renormalized
operators defined above satisfy boson like commutation relations:
\begin{eqnarray}
\left[\bar{A}_{1\mu}(k), \bar{A}^{\dagger}_{1\mu'}(k')\right ] &=&\delta_{k,k'}\delta_{\mu,\mu'},
\nonumber\\
\left[\bar{B}_{1\mu}(k), \bar{B}^{\dagger}_{1\mu'}(k')\right ] &=&\delta_{k,k'}\delta_{\mu,\mu'}f_k,
\;\;f_k=sign(C^{(2)}_k ).
\end{eqnarray}
Further, the renormalized operators are used in order to define the pnQRPA phonon operator
\begin{equation}
\Gamma^{\dagger}_{1\mu}=\sum_{k}\left[X(k)\bar{A}^{\dagger}_{1\mu}(k)+
Z(k)\bar{D}^{\dagger}_{1\mu}(k)- Y(k)\bar{A}_{1-\mu}(k)(-)^{1-\mu}  -
W(k)\bar {D}_{1-\mu}(k)(-)^{1-\mu}\right].
\end{equation}
where $\bar{D}^{\dagger}_{1\mu}(k)$ stands for either $\bar{B}^{\dagger}_{1\mu}(k)$ or
$\bar{B}_{1\mu}(k)$ depending on whether $f_k$ is equal to $+1$ or $-1$.
The amplitudes $X,Z,Y,W$ are determined by the fully FRpnQRPA equations
provided by the operator equations:
\begin{equation}
\left[H,\Gamma^{\dagger}_{1\mu}\right]=\omega \Gamma^{\dagger}_{1\mu}\;\;\left[\Gamma_{1\mu},\Gamma^{\dagger}_{1\mu'}\right]=\delta_{\mu\mu'}.
\end{equation}
The number of FRpnQRPA equations is double the number of standard pnQRPA equations.
They have to be solved at a time with the equations defining the constants
$C^{(1)}$ and $C^{(2)}$. The phonon vacuum is the ground state of the system
which has the property that the corresponding average of the quasiparticle number operator
is non-vanishing.
Solving these equations, one finds out that half of the solutions have the amplitude
X dominant while for  the remaining ones, the amplitudes Z prevail. The latter solutions have
been separately studied in Ref.\cite{Rad4}. It has been proven that such solutions describe a wobbling
motion of the isospin degrees of freedom.
\section{Restoring the gauge symmetry for the fully renormalized pnQRPA solutions}
The major component of vacuum state $|0\rangle$ is a state characterizing the even-even system
(N,Z) under consideration. Considering the inverse BV transformation for the dipole operators
$A^{\dagger},A, B^{\dagger},B$ one can easily check that  one phonon states are mixtures
of components describing the neighboring nuclei (N-1,Z+1),(N+1,Z-1),(N+1,Z+1),(N-1,Z-1).
The first two components preserve the total number of nucleons but violate
 the third isospin component $T_3$ while that remaining components violate the total number of nucleons but
 preserve $T_3$.The latter two components mentioned above are the ones which are
 responsible for the violation of Ikeda sum rule (ISR), which is valid for single beta
 decays of the (N,Z) nucleus.
 Aiming at getting a suitable structure for one phonon state, so that ISR is 
obeyed,
 it is desirable to start with linear combinations of the basic operators
 $A^{\dagger},A, B^{\dagger},B$ which excite the (N,Z) nucleus to the nuclei
(N-1,Z+1),(N+1,Z-1),(N+1,Z+1),(N-1,Z-1), respectively.
One can check that such operators are:
\begin{eqnarray}
{\cal A}^{\dag}_{1\mu}(pn)&=&U_pV_nA^{\dag}_{1\mu}(pn)+U_nV_pA_{1,-\mu}(pn)(-)^{1-\mu}+
U_pU_nB^{\dag}_{1\mu}(pn)-V_pV_nB_{1,-\mu}(pn)(-)^{1-\mu},\nonumber\\
{\cal A}_{1\mu}(pn)&=&U_pV_nA_{1\mu}(pn)+U_nV_pA^{\dag}_{1,-\mu}(pn)(-)^{1-\mu}+
U_pU_nB_{1\mu}(pn)-V_pV_nB^{\dag}_{1,-\mu}(pn)(-)^{1-\mu},
\nonumber\\
{\bf{\bf A}}^{\dag}_{1\mu}(pn)&=&U_pU_nA^{\dag}_{1\mu}(pn)-V_pV_nA_{1,-\mu}(pn)(-)^{1-\mu}-
U_pV_nB^{\dag}_{1\mu}(pn)-V_pU_nB_{1,-\mu}(pn)(-)^{1-\mu},
\nonumber\\
{\bf{\bf A}}_{1\mu}(pn)&=&U_pU_nA_{1\mu}(pn)-V_pV_nA^{\dag}_{1,-\mu}(pn)(-)^{1-\mu}-
U_pV_nB_{1\mu}(pn)-V_pU_nB^{\dag}_{1,-\mu}(pn)(-)^{1-\mu}.\nonumber\\
\end{eqnarray}
Indeed, expressed in terms of creation and annihilation particle operators the above
operators are:
\begin{eqnarray}
{\cal A}^{\dag}_{1\mu}(pn)&=&-\left[c^{\dag}_pc_{\widetilde{n}}\right]_{1\mu},\;\;
{\cal A}_{1\mu}(pn)=-\left[c^{\dag}_pc_{\widetilde{n}}\right]^{\dag}_{1\mu},
\nonumber\\
{\bf{\bf A}}^{\dag}_{1\mu}(pn)&=&\left[c^{\dag}_pc^{\dag}_n\right]_{1\mu},\;\;
{\bf{\bf A}}_{1\mu}(pn)=\left[c^{\dag}_pc^{\dag}_n\right]^{\dag}_{1\mu}.
\end{eqnarray}
In terms of the new dipole operators, the
previously introduced one body operators become:
\begin{eqnarray}
\beta^-_{\mu}(pn)&=&\frac{\hat {j}_p}{\hat {1}}\langle j_p||\sigma||j_n\rangle
{\cal A}^{\dag}_{1\mu}(pn),\;\;\beta^+_{\mu}(pn)=\frac{\hat {j}_p}{\hat {1}}\langle j_p||\sigma||j_n\rangle
{\cal A}_{1,-\mu}(pn)(-)^{\mu}, \nonumber\\
P^-_{1\mu}(pn)&=&\frac{\hat {j}_p}{\hat {1}}\langle j_p||\sigma||j_n\rangle
{\bf {\bf A}}^{\dag}_{1\mu}(pn),\;\;P^+_{\mu}(pn)=\frac{\hat {j}_p}{\hat {1}}\langle j_p||\sigma||j_n\rangle
{\bf{\bf A}}_{1,-\mu}(pn)(-)^{\mu}.
\end{eqnarray}
The model Hamiltonian can be written as:
\begin{eqnarray}
H&=&\sum E_{\tau j}a^{\dag}_{\tau jm}a_{\tau jm}+2\chi\sum\sigma_{pn;p'n'}
{\cal A}^{\dag}_{1\mu}(pn){\cal A}_{1\mu}(p'n')  -2\chi_1\sum \sigma_{pn;p'n'}
{\bf {\bf A}}^{\dag}_{1\mu}(pn){\bf {\bf A}}_{1\mu}(p'n'),\nonumber\\
\sigma_{pn;p'n'}&=&\frac{{\hat j}_p{\hat j}_{p'}}{3}\langle j_p||\sigma ||j_n\rangle
\langle j_{p'}||\sigma ||j_{n'}\rangle .
\end{eqnarray}
In order to study the harmonic modes which might be defined with the operators
${\cal A}^{\dag}_{1\mu}(pn), {\cal A}_{1\mu}(pn), {\bf {\bf A}}^{\dag}_{1\mu}(pn),
{\bf {\bf A}}_{1\mu}(pn)$ we need their mutual commutation relations. These are given in
Appendix A.
Following the prescription of the fully renormalized pnQRPA, these commutators
can be approximated as:
\begin{eqnarray}
\left[{\cal A}_{1\mu}(pn),{\cal A}^{\dag}_{a\mu '}(p'n')\right]&\approx&
\delta_{\mu,\mu'}\delta_{j_p,j_{p'}}\delta_{j_n,j_{n'}}
\left[U_p^2-U_n^2+\frac{U_n^2-V_n^2}{{\hat j}_n^2}{\hat N}_n-
\frac{U_p^2-V_p^2}{{\hat j}_p^2}{\hat N}_p\right],\nonumber\\
\left[{\bf {\bf  A}}_{1\mu}(pn),{\bf{\bf A}}^{\dag}_{a\mu '}(p'n')\right]&\approx&
\delta_{\mu,\mu'}\delta_{j_p,j_{p'}}\delta_{j_n,j_{n'}}
\left[U_p^2-V_n^2-\frac{U_n^2-V_n^2}{{\hat j}_n^2}{\hat N}_n-
\frac{U_p^2-V_p^2}{{\hat j}_p^2}{\hat N}_p\right],\nonumber\\
\left[{\cal A}_{1\mu}(pn),{\bf {\bf A}}^{\dag}_{a\mu '}(p'n')\right]&\approx&
\delta_{\mu,\mu'}\delta_{j_p,j_{p'}}\delta_{j_n,j_{n'}}U_nV_n
\left[1-2\frac{{\hat N}_n}{{\hat j}_n^2}\right],\nonumber\\
\left[{\cal A}^{\dag}_{1\mu}(pn),{\bf {\bf A}}^{\dag}_{1,-\mu '}(p'n')(-)^{1-\mu '}\right]&\approx&
\delta_{\mu,\mu'}\delta_{j_p,j_{p'}}\delta_{j_n,j_{n'}}U_pV_p
\left[1-2\frac{{\hat N}_p}{{\hat j}_p^2}\right].
\end{eqnarray}
Here, ${\hat N}_p$ and $ {\hat N}_n$ stand for the proton and neutron quasiparticle number operators, respectively:
\begin{equation}
\hat{N}_{\tau}=\sum_{jm}a^{\dagger}_{\tau jm}a_{\tau jm}; \tau =p,n.
\end{equation}
For what follows it is useful to introduce the notations:
\begin{eqnarray}
D_1(pn)&=&U_p^2-U_n^2+\frac{1}{2j_n+1}(U_n^2-V_n^2)\langle{\hat N}_n\rangle -
\frac{1}{2j_p+1}(U_p^2-V_p^2)\langle{\hat N}_p\rangle,
\nonumber\\
D_2(pn)&=&U_p^2-V_n^2-\frac{1}{2j_n+1}(U_n^2-V_n^2)\langle{\hat N}_n\rangle -
\frac{1}{2j_p+1}(U_p^2-V_p^2)\langle{\hat N}_p\rangle,
\nonumber\\
D_n&=&U_nV_n\left[1-\frac{2}{2j_n+1}\langle{\hat N}_n\rangle\right],\nonumber\\
D_p&=&U_pV_p\left[1-\frac{2}{2j_p+1}\langle{\hat N}_p\rangle\right],
\end{eqnarray}
Here the average value of an operator $\hat{O}$ on the renormalized phonon operator vacuum state, undefined for the time being, is denoted by $\langle\hat {O}\rangle$.
According to Eq. (A.1) the quantities $D_n$ and $D_p$ represent to some extent
the overlap of the ground state of the (N,Z) system with states describing
the (N+2,Z) and (N,Z+2) systems, respectively. The latter states are obtained by adding one pair of neutrons and one pair of protons to the (Z,N) ground state, respectively.
Both pairs, of neutrons and protons, have a vanishing total angular momentum.

Note that for the single particle states involved in single $\beta^-$ transitions,
$D_1(pn)$ is a positive quantity, while $D_2(pn)$ might be either positive or negative, depending on the relative values of the occupation probabilities for protons and neutrons in the states $p$ and $n$, respectively.
Let us denote by $j_c$ the critical angular momentum having the property:  
\begin{eqnarray}
 {\rm If}\;\;\epsilon_{pj}&\leq& \epsilon_{pj_c} ,\;\;\rm{then}\;\; D_2(pn)\leq 0,\nonumber\\
 {\rm If}\;\;\epsilon_{pj}& > &\epsilon_{pj_c} ,\;\;\rm{then}\;\; D_2(pn)> 0.
\end{eqnarray}
It is useful to introduce the functions:
\begin{eqnarray}
\theta_{pn}&=&\left\{\matrix{1&\;\;\rm{if}\;\;D_2(pn)>0,\cr
                           0&\;\;\rm{if}\;\;D_2(pn)\leq0, }\right.  
\nonumber\\
\bar{\theta}_{pn}&=&\left\{\matrix{1&\;\;\rm{if}\;\;D_2(pn)\leq 0, \cr
                                   0&\;\;\rm{if}\;\;D_2(pn)>0.}\right.  
\end{eqnarray}
Then the operators ${\cal A}^{\dag}_{1\mu}, {\bf {\bf A}}^{\dag}_{1\mu}$
 can be renormalized as:
\begin{eqnarray}
\bar{{\cal A}}^{\dag}_{1\mu}(pn)&=&\frac{1}{\sqrt{D_1(pn)}}{\cal A}^{\dag}_{1\mu}\;,
\nonumber\\
\bar{\bf {\bf A}}^{\dag}_{1\mu}(pn)&=&\frac{1}{\sqrt{|D_2(pn)|}}{\bf {\bf A}}^{\dag}_{1\mu}\;.
\end{eqnarray}
One can certainly introduce new operators having proper boson like normalization:
\begin{eqnarray}
\bar{{\cal B}}^{\dag}_{1\mu}(pn)&=&\theta_{pn}\bar{\bf{\bf A}}^{\dag}_{1\mu}+
\bar{\theta}_{pn}\bar {\bf {\bf A}}_{1,-\mu}(-)^{1-\mu},\nonumber\\
\bar{{\cal B}}_{1\mu}(pn)&=&\theta_{pn}\bar{\bf{\bf A}}_{1\mu}+
\bar{\theta}_{pn}\bar {\bf {\bf A}}^{\dag}_{1,-\mu}(-)^{1-\mu},\nonumber\\
\end{eqnarray}
The result is that we have two pairs of operators satisfying boson like commutation relations:
\begin{eqnarray}
\left[\bar{{\cal A}}_{1\mu}(pn) ,\bar{{\cal A}}^{\dag}_{1\mu}(p'n')\right]&=&\delta_{\mu,\mu'}
\delta_{j_p,j_{p'}} \delta_{j_n,j_{n'}},\nonumber\\
\left[\bar{{\cal B}}_{1\mu}(pn) ,\bar{{\cal B}}^{\dag}_{1\mu}(p'n')\right]&=&\delta_{\mu,\mu'}
\delta_{j_p,j_{p'}} \delta_{j_n,j_{n'}}.
\end{eqnarray}
In this Section  we suppose that $D_p$ and $D_n$ are negligible small. Also, we ignore
the contribution of the pairing interaction  coupling the equations of motion associated to the operators
${\cal A}^{\dag}_{1\mu}$ and ${\bf {\bf A}}^{\dag}_{1\mu}$.
Under these circumstances  the two boson operators are independent from each other,
\begin{equation}
\left[\bar{{\cal B}}_{1\mu}(pn) ,\bar{{\cal A}}^{\dag}_{1\mu}(p'n')\right]=0,
\end{equation}

and consequently the equations of motion for the quoted operators are decoupled .
\begin{eqnarray}
\left[H,\bar{{\cal A}}^{\dag}_{1\mu}(pn)\right]&=&
\left[E_p(U_p^2-V_p^2)+E_n(V_n^2-U_n^2)\right] \bar{{\cal A}}^{\dag}_{1\mu}(pn)
+2\chi\sum\sigma^{(1)}_{pn;p_1n_1}\bar{{\cal A}}^{\dag}_{1\mu}(p_1n_1),\\
\left[H,\bar{{\bf{\bf A}}}^{\dag}_{1\mu}(pn)\right]&=&
\left[E_p(U_p^2-V_p^2)+E_n(U_n^2-V_n^2)\right]\bar{{\bf{\bf A}}}^{\dag}_{1\mu}(pn)
-2\chi_1\varphi_{pn}\sum\sigma^{(2)}_{pn;p_1n_1}\bar{{\bf{\bf A}}}^{\dag}_{1\mu}(p_1n_1).
\nonumber
\end{eqnarray}
Here the following notations have been used:
\begin{eqnarray}
\sigma^{(k)}_{pn;p_1n_1}&=&\frac{{\hat j}_p}{{\hat 1}}\langle p||\sigma||n\rangle
|D_k(p,n)|^{1/2}\frac{{\hat j}_{p_1}}{{\hat 1}}\langle p_1||\sigma||n_1\rangle
|D_k(p_1,n_1)|^{1/2},\;\;k=1,2\nonumber\\
\varphi_{pn}&=&\frac{D_2(pn)}{|D_2(pn)|}.
\end{eqnarray}
We look for the linear combination operators
\begin{equation}
\Gamma^{\dag}_{1\mu}=\sum_{p,n}X(pn)\bar{{\cal A}}^{\dag}_{1\mu}(pn),\;\;
{\cal G}^{\dag}_{1\mu}=\sum_{p,n}{\cal X}(pn)\bar{{\bf {\bf A}}}^{\dag}_{1\mu}(pn)
\end{equation}
with the properties:
\begin{eqnarray}
\left[H,\Gamma^{\dag}_{1\mu} \right]&=&\omega \Gamma^{\dag}_{1\mu}, \;\;\left[\Gamma_{1\mu},\Gamma^{\dag}_{1\mu'}\right]=\delta_{\mu,\mu'}, \nonumber\\
\left[H,{\cal G}^{\dag}_{1\mu} \right]&=&\Omega {\cal G}^{\dag}_{1\mu},\;\;\left[{\cal G}_{1\mu},{\cal G}^{\dag}_{1\mu'}\right]=\delta_{\mu,\mu'}.
\label{TDEQ}
\end{eqnarray}
These equations provide  homogeneous systems of linear equations for the amplitudes
$X$ and ${\cal X}$, respectively. The compatibility condition yields for $\omega$ and $\Omega$ the dispersion equations:
\begin{eqnarray}
2\chi\sum_{p,n}\frac{\frac{2j_p+1}{3}\left(\langle p||\sigma ||n\rangle \right)^2
D_1(pn)}{E_{p}(U_{p}^2-V_p^2)+E_n(V_n^2-U_n^2)-\omega}+1&=&0,\nonumber\\
2\chi_1\sum_{p,n}\frac{\frac{2j_p+1}{3}\left(\langle p||\sigma ||n\rangle \right)^2
D_2(pn)}{E_{p}(U_{p}^2-V_p^2)+E_n(U_n^2-V_n^2)-\Omega}-1&=&0.
\label{DispEq}
\end{eqnarray}
The distinct feature of these equations consists of that the poles of 
the functions involved in the l.h.s. of Eq.(\ref{DispEq}) are simple. Thus, while in the standard pnQRPA the dispersion equation depends on the squared energy,
 here the dependence is simply on energy. The important consequence is that the equations derived above do not have vanishing solutions.

The states created by acting with the phonon operators $\Gamma^{\dagger}$ and 
${\cal G}^{\dagger}$ on the ground state of an (Z,N) system are dipole states $1^{+}$ in the neighboring odd-odd (Z+1,N-1) and (Z+1,N+1) nuclei. While the first state has a particle-hole ($ph$) character, the second one is a particle-particle ($pp$) like excitation.
The phonon amplitudes have the expressions:
\begin{eqnarray}
X(pn)&=&- 2\chi\frac{\frac{\hat{j}_p}{\hat{1}}\langle p||\sigma ||n\rangle
D^{1/2}_1(pn)}{E_{p}(U_{p}^2-V_p^2)+E_n(V_n^2-U_n^2)-\omega}S, \nonumber\\
{\cal X}(pn)&=&2\chi_1\varphi_{pn}\frac{\frac{\hat{j}_p}{\hat{1}}\langle p||\sigma ||n\rangle
|D_2(pn)|^{1/2}}{E_{p}(U_{p}^2-V_p^2)+E_n(U_n^2-V_n^2)-\Omega}{\cal S}.
\end{eqnarray}
The factors $S$ and ${\cal S}$ are determined by the second equation from (\ref{TDEQ}),
requiring that the phonon operators are normalized to unity. The final results are:
\begin{eqnarray}
S^{-1}&=& 2\chi\left[\sum_{p,n}\frac{\frac{2j_p+1}{3}\left(\langle p||\sigma ||n\rangle \right)^2
D_1(pn)}{\left(E_{p}(U_{p}^2-V_p^2)+E_n(V_n^2-U_n^2)-\omega\right)^2} \right]^{1/2},
\nonumber\\
{\cal S}^{-1}&=& 2\chi_1\left[\sum_{p,n}\frac{\frac{2j_p+1}{3}\left(\langle p||\sigma ||n\rangle \right)^2
D_2(pn)}{\left(E_{p}(U_{p}^2-V_p^2)+E_n(U_n^2-V_n^2)-\Omega\right)^2} \right]^{1/2}.
\end{eqnarray}
Note that both the dispersion equations and the phonon amplitudes depend on
the renormalization factors $D$,  which at
their turn depend on the quoted amplitudes. Therefore, the FRpnQRPA equations and
the defining equation (3.7) should be solved at a time. In order to do that we have
to provide a recipe for how to calculate the averages for the quasiparticle number operators
involved in Eq. (3.7). In what follows, we shall describe separately the results for the
phonon operators $\Gamma^{\dagger}$ and ${\cal G}^{\dagger}$. As we proceeded in Ref.\cite{Rad3}, we seek
for a boson representation of the quasiparticle number operators, determined so
that
their commutation relations with the phonon operators are preserved.
\begin{eqnarray}
{\hat N}_p&=&\sum_{k,\mu}C_k(p)\Gamma_{1\mu}(k)\Gamma^{\dagger}_{1\mu}(k),\nonumber\\
{\hat N}_n&=&\sum_{k,\mu}C_k(n)\Gamma_{1\mu}(k)\Gamma^{\dagger}_{1\mu}(k).
\end{eqnarray}
The expansion coefficients have the following expressions:
\begin{equation}
C_k(\tau)=-\langle 0|\left[\left[ {\hat N}_{\tau},\Gamma_{1\mu}(k)\right],\Gamma^{\dagger}_{1\mu}(k)\right]|0\rangle,
\;\;\tau=p,n.
\end{equation}
Here $|0\rangle$ stands for the FRpnQRPA vacuum.
Calculating the commutators involved in the above equations, one finds:
\begin{eqnarray}
\langle {\hat N}_p\rangle &=&\sum_{k}C_k(p)
\nonumber\\
&=&\sum_{k,n}\frac{|X_k(pn)|^2}{D_1(pn)}
\left(U_p^2V_n^2+U_n^2V_p^2+\frac{\langle {\hat N}_n\rangle}{{\hat j}_n^2}
(U_n^2-V_n^2)(U_p^2-V_p^2)
-\frac{\langle {\hat N}_p\rangle}{{\hat j}_p^2}\right),\nonumber\\
\langle {\hat N}_n\rangle &=&\sum_{k}C_k(n)
\nonumber\\
&=&-\sum_{k,p}\frac{|X_k(pn)|^2}{D_1(pn)}
\left(U_p^2V_n^2+U_n^2V_p^2+\frac{\langle {\hat N}_n\rangle}{{\hat j}_n^2}(U_n^2-V_n^2)(U_p^2-V_p^2)
-\frac{\langle {\hat N}_p\rangle}{{\hat j}_p^2}\right).
\end{eqnarray}
These equations allow to express the averages $\langle {\hat N}_p \rangle$
and $\langle {\hat N}_n \rangle$  as functions of $D_1(pn)$. Inserting the results in
(3.7), a set of equations for the renormalization factors is obtained.
Since the averaged quasiparticle numbers depend on the $X$ amplitudes, so do the
renormalization factors $D_1(pn)$.

As for  the deuteron like phonon operator the results are:
\begin{eqnarray}
{\hat N}_p&=&\sum_{k,\mu}F_k(p){\cal G}_{1\mu}(k){\cal G}^{\dagger}_{1\mu}(k),
\nonumber\\
{\hat N}_n&=&\sum_{k,\mu}F_k(n){\cal G}_{1\mu}(k){\cal G}^{\dagger}_{1\mu}(k),
\end{eqnarray}
where
\begin{equation}
F_k(\tau)=-\langle 0|\left[\left[ {\hat N}_{\tau},{\cal G}_{1\mu}(k)\right],{\cal G}^{\dagger}_{1\mu}(k)\right]|0\rangle,
\;\;\tau=p,n.
\end{equation}
Using the final results for the coefficients $F$, one obtains:
\begin{eqnarray}
\langle {\hat N}_p\rangle &=&\sum_{k}F_k(p)
\nonumber\\
&=&\sum_{k,n}\frac{|{\cal X}_k(pn)|^2\varphi_{pn}}{D_2(pn)}
\left(U_p^2U_n^2+V_p^2V_n^2-\frac{\langle {\hat N}_n\rangle}{{\hat j}_n^2}
(U_n^2-V_n^2)(U_p^2-V_p^2)
-\frac{\langle {\hat N}_p\rangle}{{\hat j}_p^2}\right),\nonumber\\
\langle {\hat N}_n\rangle &=&\sum_{k}F_k(n)
\\
&=&-\sum_{k,p}\frac{|{\cal X}_k(pn)|^2\varphi_{pn}}{D_2(pn)}
\left(U_p^2U_n^2+V_p^2V_n^2-\frac{\langle {\hat N}_n\rangle}{{\hat j}_n^2}(U_n^2-V_n^2)(U_p^2-V_p^2)
-\frac{\langle {\hat N}_p\rangle}{{\hat j}_p^2}\right).\nonumber
\label{NpNn}
\end{eqnarray}
These can be viewed as a system of nonlinear equations for $\langle \hat {N}_p\rangle $ and 
$\langle \hat {N}_n\rangle $ for a given set of
amplitudes ${\cal X}$. Solving these equations and inserting the solutions in Eq.(3.7)
one obtains the values of $D_2(pn)$. An easier way would be to express
$\langle \hat {N}_p\rangle $ and $\langle \hat {N}_n\rangle $ as function of $D_2(pn)$ and then Eq.(3.7) becomes a
nonlinear equation for $D_2(pn)$ that should be solved at a time with the equations
(3.18) and (3.19).

Consider, again, the equations associated to the operators ${\cal A}^{\dagger}$ and
 ${\cal A}$. The quantities $D_1(pn)$ should fulfill an additional consistency
 equation caused by the fact that the phonon operator commutes with the quasiparticle total number operator.
Requiring that also $T_3$ is preserved in the average, one obtains that the
numbers of protons and neutrons are
separately preserved.
The average values of the proton and neutron number operators corresponding to 
the renormalized pnQRPA ground state have the expressions:
\begin{eqnarray}
Z=\sum_{p}\hat{j}_p^2(V_p^2+\frac{U_p^2-V_p^2}{\hat{j}_p^2}\langle \hat{N}_p\rangle),\nonumber\\ 
N=\sum_{p}\hat{j}_n^2(V_n^2+\frac{U_n^2-V_n^2}{\hat{j}_n^2}\langle \hat{N}_n\rangle),
\label{ZandN}
\end{eqnarray}
These equations suggest that in order to have a pnQRPA ground state with non-vanishing
 quasiparticle numbers and with good proton and neutron numbers it is necessary to
 renormalize the BCS equations for chemical potentials. In this way the BCS and fully
 FRpnQRPA equations are coupled to each other. It should be mentioned that the
 above equations have
 been derived by assuming that the averages of the quasiparticle pair operators are vanishing. There is still an alternative option, namely to keep the standard BCS equations as they are
 but enforce the consistency restrictions:
 \begin{eqnarray}
 \sum_{p} U_pV_p\hat{j}_p\langle \left([a^{\dagger}_pa^{\dagger}_{\tilde {p}}]_{00}
 +[a_{\tilde {p}}a_p]_{00}\right)\rangle &=&\sum_{p}(V_p^2-U_p^2)
 \langle {\hat N}_p\rangle,
 \nonumber\\
\sum_{n}U_nV_n\hat{j}_n\langle \left([a^{\dagger}_na^{\dagger}_{\tilde {n}}]_{00}
 +[a_{\tilde {n}}a_n]_{00}\right)\rangle &=&\sum_{n}(V_n^2-U_n^2)\langle {\hat N}_n\rangle,
\end{eqnarray}
As we shall see a bit later, following the second path sketched before, the
ISR is violated.

Despite the fact the  phonon operator
defined before is of a Tamm-Dancoff type, it might be viewed
as a FRpnQRPA phonon operator acting in the extended $4N_s$ dimensional space. Therefore, the present
results reproduce a set of $N_s$ solutions, with $N_s$ denoting the number of proton-neutron states (p,n) which can couple at an angular momentum equal to unity, out of 2$N_s$  positive solutions provided by FRpnQRPA.

Indeed, denoting by
\begin{eqnarray}
D_A(pn)&=&1-\frac{1}{2j_n+1}\langle {\hat N}_n\rangle-\frac{1}{2j_p+1}\langle {\hat N}_p\rangle,\nonumber\\
D_B(pn)&=&\frac{1}{2j_n+1}\langle {\hat N}_n\rangle-\frac{1}{2j_p+1}\langle {\hat N}_p\rangle,
\end{eqnarray}
the phonon operator can be expressed in terms of the renormalized operators:
\begin{equation}
\bar{A}^{\dag}_{1\mu}(pn)=\frac{1}{\sqrt{D_A(pn)}}A^{\dag}_{1\mu}(pn),\;
\bar{B}^{\dag}_{1\mu}(pn)=\frac{1}{\sqrt{|D_B(pn)|}}B^{\dag}_{1\mu}(pn).
\end{equation}
\begin{eqnarray}
\Gamma^{\dag}_{1\mu}
&=&\sum_{p,n} X(pn)D^{-1/2}_1(pn)\left(U_pV_nD^{1/2}_A\bar{A}^{\dag}_{1\mu}(pn)
+U_nV_pD^{1/2}_A\bar{A}_{1,-\mu}(pn)(-)^{1-\mu}\right.\\
&+&\left.U_pU_n|D_B|^{1/2}\bar{B}^{\dag}_{1\mu}(pn)
-V_nV_p|D_B|^{1/2}\bar{B}_{1,-\mu}(pn)(-)^{1-\mu}\right)\nonumber\\
&\equiv&\sum \left [{\bf X}(pn)\bar{A}^{\dag}_{1\mu}(pn)-{\bf Y}(pn)
\bar{A}_{1,-\mu}(pn)(-)^{1-\mu}+{\bf Z}(pn)\bar{B}^{\dag}_{1\mu}(pn)
-{\bf W}(pn)\bar{B}_{1,-\mu}(pn)(-)^{1-\mu}\right].   \nonumber
\label {Gaext}
\end{eqnarray}
The norm for the extended phonon is:
\begin{eqnarray}
&&\sum_{p,n}\left({\bf X}^2(pn)-{\bf Y}^2(pn)+{\bf Z}^2(pn)-{\bf W}^2(pn)\right)
\nonumber\\
&=&                                                                             \sum_{p,n}X^2(pn)\frac{1}{D_1(pn)}\left[\left(U_p^2V_n^2-U_n^2V_p^2\right)
\left(1-\frac{\langle
{\hat N}_n\rangle}{{\hat j}^2_n}- \frac{\langle
{\hat N}_p\rangle}{{\hat j}^2_p}\right)+\left(U_p^2U_n^2-V_p^2V_n^2\right)\left(
\frac{\langle
{\hat N}_n\rangle}{{\hat j}^2_n}- \frac{\langle
{\hat N}_p\rangle}{{\hat j}^2_p}\right) \right]\nonumber\\
&=&\sum_{pn} X^2(pn)\frac{1}{D_1(pn)}\left[U_p^2-V_n^2+\left(U_n^2-V_n^2\right)
\frac{\langle
{\hat N}_n\rangle}{{\hat j}^2_n}-\left(U_p^2-V_p^2\right)
\frac{\langle
{\hat N}_p\rangle}{{\hat j}^2_p}\right]=\sum_{p,n}X^2(pn)=1.
\end{eqnarray}

Thus, the Tamm-Dancoff phonon operator can be embedded in the operator
space acting on the fully renormalized pnQRPA states. Reciprocally, Eq.(\ref{Gaext}) tells us how to obtain the Tamm-Dancoff phonon amplitudes as functions of the fully renormalized pnQRPA phonon amplitudes.

Note that, although the operator ${\cal A}^{\dagger}$ is a particle
hole operator in  particle representation, the Tamm-Dancoff equations accounts
for quasiparticle correlations and therefore describe the 
system's small oscillations
around a static BCS ground state.

Let us check now whether the Ikeda sum rule is obeyed by the ground state corresponding to the Tamm-Dancoff phonon $\Gamma^{\dagger}$.
ISR is generated by the identity: 

\begin{equation}
\sum_{\mu}\left[\beta^{\dag}_{-\mu},\beta^-_{\mu}\right](-)^{\mu}=3\left[{\hat {\cal N}}_n-{\hat {\cal N}}_p\right],
\end{equation}
where $\hat{{\cal N}}_{\tau}$ ($\tau=p,n$) denotes the $\tau$-particle number operator. Averaging this equation on the renormalized ground state and then inserting between the two operators
 $\beta^{\pm}$ the unity operator, one obtains:
\begin{eqnarray}
S_I&=&\sum_{\mu,k}\langle0|\beta^{\dag}_{-\mu}|1_k\rangle\langle 1_k|\beta^-_{\mu} |0\rangle
\nonumber\\
&=&
\sum_{p,n;p'n';\mu}\frac{{\hat j}_p}{{\hat 1}}\langle j_p||\sigma ||j_n\rangle\langle 0|{\cal A}_{\mu}
(pn)|1_k\rangle \langle 1_k|{\cal A}^{\dag}_{1\mu}(p'n')|0\rangle
\frac{{\hat j}_{p'}}{{\hat 1}}\langle j_{p'}||\sigma ||j_{n'}\rangle
\nonumber\\
&=&\sum_{p,n;p'n';k} \frac{{\hat j}_p}{{\hat 1}}\langle j_p||\sigma ||j_n\rangle
\frac{{\hat j}_{p'}}{{\hat 1}}\langle j_{p'}||\sigma ||j_{n'}\rangle
X_k(pn)X_k(p'n').3
\nonumber\\
&=&\sum_{p,n}(2j_p+1)\left(\langle j_p||\sigma ||j_n\rangle\right)^2D_1(pn).
\end{eqnarray}
Here the index k is an ordering label for the roots of the dispersion equation (3.18).
Note that $D_1(pn)$ is a sum of terms depending exclusively on 
$p$ and $n$ respectively, and moreover
\begin{eqnarray}
\sum_p|\langle j_p||\sigma||j_n\rangle |^2&=&3,\nonumber\\
\sum_n|\langle j_n||\sigma||j_p\rangle |^2&=&3.
\end{eqnarray}
Taking now into account the result from Eq.(\ref{ZandN}) one obtains:
\begin{equation}
S_I=3(N-Z).
\end{equation}
We may conclude that within the present formalism the Ikeda sum rule is satisfied.
A peculiar feature of the present result is that ISR is satisfied although
the strength of the $\beta^+$ transition is vanishing.
Therefore, projecting out the components of good particle total number from the
phonon operator, one obtains a phonon operator which is formally of Tamm-Dancoff
type and describes a subset of $pn$ excitations obtainable due to the
unprojected operator. Moreover, ignoring the terms coupling the equations
of motion of the particle total number conserving operators with those of
operators non-conserving the particle total number, the two body $pp$ interaction is ruled out.
We have seen that assuming a consistent decoupling scheme, two independent excitations
appear, one describing the (N-1,Z+1) system while the other one the (N+1,Z+1) nucleus. The first excitations are suitable for describing the rates of the single $\beta^-$ process while the other type of states are deuteron like states and might be populated in a transfer reaction process. Similar formalism can be
immediately written down for the states in the neighboring 
(N+1,Z-1), (N-1,Z-1) nuclei obtained in a $\beta^+$ process and by removing a dipole deuteron
from the mother nucleus, respectively.
\section{A possible improvement of the FRpnQRPA approach}
In the previous section we have assumed that the commutators of the operators
${\cal A}^{\dagger}$ and ${\bf A}^{\dagger}$ are negligible small.
Let us now investigate whether it is possible to define a linear combination
\begin{equation}
C^{\dagger}_{1\mu}=X(k){\bf {\bf A}}^{\dagger}_{1\mu}(k)+Y(k)
{\bf {\bf A}}_{1\mu}(k)(-)^{1-\mu},
\end{equation}
which commutes with both  ${\cal A}^{\dagger}_{1\mu}$ and ${\cal A}$, irrespective of $D_p$ and $D_n$ magnitudes.
One  finds out that a necessary condition is that $D_p(k)=D_n(k)$ which results in having either $X(k)=Y(k)$ or $X(k)=-Y(k)$. A similar result is obtained
for a linear combination of ${\cal A}^{\dagger}$ and ${\cal A}$ required to 
commute with ${\bf {\bf A}}^{\dagger}$ and  ${\bf {\bf A}}$.
One may conclude that, rigorously speaking, it is not possible to define  non-spurious pnQRPA phonon
operators as linear superposition of ${\cal A}^{\dagger}$ and ${\cal A}$
which are fully decoupled of the phonon operators built up with 
${\bf {\bf A}}^{\dagger}$ and  ${\bf {\bf A}}$. In this context one may state that the linear combination of ${\cal A}^{\dagger}$ and ${\cal A}$ used
 as phonon operator, in Ref.\cite{Ro}, which commutes with the particle total number operator is not justified.

However, it can be checked that there is  a linear combination operator
\begin{equation}
{\cal R}^{\dagger}_{1\mu}(pn)=a{\bf{\bf A}}^{\dagger}_{1\mu}(pn)+b{\bf {\bf A}}_{1,-\mu}(-)^{1-\mu}(pn)+c{\cal A}^{\dagger}_{1\mu}(pn)+
z{\cal A}_{1,-\mu}(-)^{1-\mu}(pn).\end{equation}
corresponding to a specific set of coefficients
\begin{eqnarray}
a&=&\frac{U_n^2-V_n^2}{2U_nV_n(U_p^2-U_n^2)},\;\;
b=\frac{U_p^2-V_p^2}{2U_pV_p(U_p^2-U_n^2)},\;\;c=\frac{1}{U_p^2-U_n^2},
\nonumber\\
z&=&\frac{1}{2D_1(U_p^2-U_n^2)}\left(\frac{U_n^2-V_n^2}{U_nV_n}D_n-\frac{U_p^2-V_p^2}{U_pV_p}D_p\right),
\end{eqnarray}
which commutes with ${\cal A}^{\dagger}_{1\mu}$ and ${\cal A}_{1,-\mu}(-)^{1-\mu}$
at a time.
Due to this property, it is worth building up a phonon operator out of the independent operators 
$({\cal A}^{\dagger}, {\cal A}_{1-\mu}(-)^{1-\mu})$;$({\cal R}^{\dagger}_{1\mu},{\cal R}_{1,-\mu}(-)^{1-\mu})$.
Note the the new proton-neutron operators can also be renormalized, 
\begin{equation}
\bar{{\cal R}}^{\dagger}_{1\mu}(pn)=\frac{1}{\sqrt{|D_3(pn)|}}{\cal R}^{\dagger}_{1\mu}(pn)\;\;\bar{{\cal R}}_{1,-\mu}(-)^{1-\mu}(pn)=\frac{1}{\sqrt{|D_3(pn)|}}{\cal R}_{1,-\mu}(pn)(-)^{1-\mu}.
\end{equation}
where
\begin{equation}
D_3(pn)=(c^2-z^2)D_1(pn)+(a^2-b^2)D_2(pn)+2(az-bc)D_p+2(ac-bz)D_n,
\end{equation}
so that the new operators satisfy boson like commutation relations:
\begin{equation}
\left[{\bar {\cal R}}_{1\mu}(pn),{\bar {\cal R}}^{\dagger}_{1\mu'}(p'n')\right]=\delta_{p,p'}\delta_{n,n'}\delta_{\mu,\mu'}.
\end{equation}

After some laborious, otherwise elementary, calculations one finds the equations of motion for the renormalized operators
${\cal A}^{\dagger}, {\cal A}_{1-\mu}(-)^{1-\mu}, {\cal R}^{\dagger}_{a\mu},{\cal R}_{1,-\mu}(-)^{1-\mu}$:
\begin{eqnarray}
\left[H,{\bar {\cal A}}^{\dagger}_{1\mu}(p_1n_1)\right]&=&\sum_{pn}\left[
T_{11}(p_1n_1;pn){\bar {\cal A}}^{\dagger}_{1\mu}(pn)+
T_{12}(p_1n_1;pn){\bar {\cal A}}_{1,-\mu}(pn)(-)^{1-\mu}+\right.
\nonumber\\
& &\left.T_{13}(p_1n_1;pn){\bar {\cal R}}^{\dagger}_{1\mu}(pn)+
T_{14}(p_1n_1;pn){\bar {\cal R}}_{1,-\mu}(pn)(-)^{1-\mu}\right],
\nonumber\\
\left[H,{\bar {\cal A}}_{1,-\mu}(p_1n_1)(-)^{1-\mu}\right]&=&\sum_{pn}\left[
T_{21}(p_1n_1;pn){\bar {\cal A}}^{\dagger}_{1\mu}(pn)+
T_{22}(p_1n_1;pn){\bar {\cal A}}_{1,-\mu}(pn)(-)^{1-\mu}+\right.
\nonumber\\
&&\left.T_{23}(p_1n_1;pn){\bar {\cal R}}^{\dagger}_{1\mu}(pn)+
T_{24}(p_1n_1;pn){\bar {\cal R}}_{1,-\mu}(pn)(-)^{1-\mu}\right],
\nonumber\\
\left[H,{\bar {\cal R}}^{\dagger}_{1\mu}(p_1n_1)\right]&=&\sum_{pn}\left[
T_{31}(p_1n_1;pn){\bar {\cal A}}^{\dagger}_{1\mu}(pn)+
T_{32}(p_1n_1;pn){\bar {\cal A}}_{1,-\mu}(pn)(-)^{1-\mu}+\right.
\nonumber\\
&&\left.T_{33}(p_1n_1;pn){\bar {\cal R}}^{\dagger}_{1\mu}(pn)+
T_{34}(p_1n_1;pn){\bar {\cal R}}_{1,-\mu}(pn)(-)^{1-\mu}\right],
\nonumber\\
\left[H,{\bar {\cal R}}_{1\mu}(p_1n_1)(-)^{1-\mu}\right]&=&\sum_{pn}\left[
T_{41}(p_1n_1;pn){\bar {\cal A}}^{\dagger}_{1\mu}(pn)+
T_{42}(p_1n_1;pn){\bar {\cal A}}_{1,-\mu}(pn)(-)^{1-\mu}+\right.
\nonumber\\
&&\left.T_{43}(p_1n_1;pn){\bar {\cal R}}^{\dagger}_{1\mu}(pn)+
T_{44}(p_1n_1;pn){\bar {\cal R}}_{1,-\mu}(pn)(-)^{1-\mu}\right].
\label{EQMOT}
\end{eqnarray}
Analytical expressions for the coefficients $T_{ik}$ involved in Eq.(\ref{EQMOT}) are given in Appendix B.
At this stage a phonon operator can be defined:

\begin{equation}
\Gamma^{\dagger}=\sum_{k=(p,n)}\left[X(k)\bar{{\cal A}}^{\dagger}_{1,\mu}(k)
+Z(k)\bar{R}^{\dagger}_{1\mu}(k)-Y(k)\bar{{\cal A}}_{1,-\mu}(k)(-)^{1-\mu})-W(k)
\bar{R}_{1,-\mu}(k)(-)^{1-\mu}\right],
\end{equation}
by requiring that the following equations are obeyed:
\begin{equation}
\left[H,\Gamma^{\dagger}_{1\mu}\right]=\omega\Gamma^{\dagger}_{1\mu}
,\;\;\left[\Gamma_{1\mu},\Gamma^{\dagger}_{1\mu'}\right]=\delta_{\mu\mu'}.
\label{HaGa}
\end{equation}
The above operator  equations provide a system of linear and homogeneous equations for the phonon amplitudes:
\begin{eqnarray}
\left(\matrix{{\cal A}&{\cal B}\cr
               -{\cal B}^*&-{\cal A}^*}\right)\left(\matrix{X\cr Z\cr Y \cr W}
\right)=\omega\left(\matrix{X\cr Z\cr Y \cr W}\right),
\end{eqnarray}
where the sub-matrices ${\cal A}$ and ${\cal B}$ have the following expressions in terms of the matrices T involved in the equations of motion written above:
\begin{eqnarray}
{\cal A}=\left(\matrix{{\bar T}_{11}&{\bar T}_{31}
\cr
{\bar T}_{13}&{\bar T}_{33}}\right),
\end{eqnarray},
\begin{eqnarray}
{\cal B}=\left(\matrix{{\bar T}_{12}^*&{\bar T}_{32}^*\cr
{\bar T}_{14}^*&{\bar T}_{34}^*}\right).
\end{eqnarray}
Here ${\bar T}_{ik}$ denotes the transposed matrix of $T_{ik}$:
\begin{equation}
{\bar T}_{ik}(p_1n_1,pn)=T_{ik}(pn,p_1n_1).
\end{equation} 
The second equation (\ref{HaGa}) yields the normalization equation for  the 
phonon amplitudes:
\begin{equation}
\sum_{k}\left[|X(k)|^2+|Z(k)|^2-|Y(k)|^2-|W(k)|^2\right]=1.
\end{equation}
The results for the normalization factor $D_3(pn)$ are as follows.
First one determines the boson expansions:
\begin{eqnarray}
{\hat N}_p&=&\sum_{k,\mu}G_k(p)\Gamma_{1\mu}(k)\Gamma^{\dagger}_{1\mu}(k),
\nonumber\\
{\hat N}_n&=&\sum_{k,\mu}G_k(n)\Gamma_{1\mu}(k)\Gamma^{\dagger}_{1\mu}(k),
\end{eqnarray}
with
\begin{equation}
G_k(\tau)=-\langle 0|\left[\left[ {\hat N}_{\tau},\Gamma_{1\mu}(k)\right],
\Gamma^{\dagger}_{1\mu}(k)\right]|0\rangle,
\;\;\tau=p,n.
\end{equation}
The final result for the average values are:
\begin{eqnarray}
\langle {\hat N}_p\rangle&=&\sum_{n,k}\left[\left(\bar{X}_k^2(pn)+
\bar{Y}_k^2(pn)\right)+\frac{\langle {\hat N}_n\rangle}{\hat{j}_n^2}
\left(\bar{Z}_k^2(pn)+
\bar{W}_k^2(pn)-\bar{X}_k^2(pn)-
\bar{Y}_k^2(pn)\right)\right.\nonumber\\
&-&\left.\frac{\langle {\hat N}_p\rangle}{\hat{j}_p^2}
\left(\bar{Z}_k^2(pn)+
\bar{W}_k^2(pn)+\bar{X}_k^2(pn)+
\bar{Y}_k^2(pn)\right)\right],\nonumber\\
\langle {\hat N}_n\rangle&=&-\sum_{p,k}\left[\left(\bar{X}_k^2(pn)+
\bar{Y}_k^2(pn)\right)+\frac{\langle {\hat N}_n\rangle}{\hat{j}_n^2}
\left(\bar{Z}_k^2(pn)+
\bar{W}_k^2(pn)-\bar{X}_k^2(pn)-
\bar{Y}_k^2(pn)\right)\right.\nonumber\\
&-&\left.\frac{\langle {\hat N}_p\rangle}{\hat{j}_p^2}
\left(\bar{Z}_k^2(pn)+
\bar{W}_k^2(pn)+\bar{X}_k^2(pn)+
\bar{Y}_k^2(pn)\right)\right].
\label{NpNn1}
\end{eqnarray}
$\bar{X},\bar{Y},\bar{Z},\bar{W}$ are the amplitudes
of the phonon operator $\Gamma^{\dagger}$ when it is expressed in terms of the primary operators
$A^{\dagger},A,B^{\dagger},B$. Their analytical expressions are given in Appendix C.
From Eq. (\ref{NpNn1}) one determines the averages $\langle \hat{N}_p \rangle$ and
$\langle \hat{N}_n \rangle$ and then by means of (3.7) and (4.5) the expressions
 of $D_1(pn)$ and $D_3(pn)$ are readily obtained.

Following a similar procedure as in the previous section, the Ikeda sum rule can be easily calculated. The result is:
\begin{equation}
S_I=\sum_{pn,p'n';k}\hat{j}_p\hat{j}_{p'}\langle j_p||\sigma||j_n\rangle
\langle j_{p'}||\sigma||j_{n'}\rangle D^{1/2}_1(pn)D^{1/2}_1(p'n')
\left(X_k(pn)X_k(p'n')-Y_k(pn)Y_k(p'n')\right),
\end{equation}
where the low index k distinguishes the FRpnQRPA solutions.

If the components $Z$ and $W$ of the FRpnQRPA mode are ignored then the summation over $k$ in the above equation would give $\delta_{pp'}\delta_{nn'}$
and
\begin{equation}
S_I=\sum_{p}(2j_p+1)\left(\langle j_p||\sigma||j_n\rangle\right)^2D_1(pn).
\end{equation} 
From this stage on one may follow the path sketched in the previous Section to 
prove that
\begin{equation}
S_I=3(N-Z).
\end{equation}
However such a picture is not fully consistent. Indeed, ignoring the
amplitudes
$Z$ and $W$ means to neglect the $T_{13}, T_{14},T_{23},T_{24}$ terms from the
equations of motion. To be consistent, one has to neglect also the $\chi_1$
terms form $T_{11}$ and $T_{12}$ which would lead to a phonon operator to
which the two body $pp$ interaction does not contribute at all. Moreover, recall the fact that the commutator $[{\cal A},{\bf {\bf A}}]$ was neglected arguing that factors like
$U_pV_p$ and $U_nV_n$ are small. Based on similar arguments, the  $UV$ terms from $T_{11}$ and $T_{12}$ have also to be left out. Then one arrives at the Tamm-Dancoff structure for the phonon operator as we obtained in the previous Section.

Coming back to Ikeda sum rule, it is obvious that the summation over $k$ of
$\left(X_k(pn)X_k(p'n')-Y_k(pn)Y_k(p'n')\right)$ is an under-unity  but close
to unity quantity for $p=p'$ and $n=n'$, and close to zero when $(p',n')\neq (pn)$.
It results that ISR is slightly underestimated by the fully renormalized and consistent pnQRPA approach.

\section{Summary and Conclusions}
In the previous sections we analyzed two distinct ways of rearranging the
two quasiparticle and quasiparticle density dipole operators inside the FRpnQRPA phonon operators, determined by the decoupling hypothesis adopted.
In the first scheme we supposed that the operators ${\cal A}^{\dagger}$ and ${\cal A}$ commuting with the particle total number operator, are fully decoupled from those which miss this property, i.e. ${\bf{\bf A}}^{\dagger},{\bf {\bf A}}$.
Also, we neglected the contribution to the equations of motions for the above quoted operators, generated by the pairing interaction which are comparable small with
the small terms mentioned before. The result is that one obtains two decoupled sets of equations of motion, both of Tamm-Dancoff type. One describes a particle-hole dipole excitation in the (N-1,Z+1) nucleus while the other one is associated to a deuteron dipole excitation in the neighboring (N+1,Z+1) nucleus.
For the first case the ISR is fully satisfied provided the BCS equations are
renormalized as well \cite{Jol,Kri,Boby}.
It is proved that, rigorously speaking,  to define a phonon operator mixing both ${\cal A}^{\dagger}$ and ${\cal A}$  and keeping at a time  the decoupling picture of the operators breaking the gauge symmetry associated with the particle 
total number, is not possible.
However it is possible to define  a linear combination of 
${\cal A}^{\dagger}$, ${\cal A}$,
${\bf {\bf A}}^{\dagger}$ and  ${\bf {\bf A}}$ denoted by $ {\cal R}^{\dagger}$ which commutes simultaneously with ${\cal A}^{\dagger}$ and ${\cal A}$. 
Finally, the phonon operator is built up based on the independent operators
${\cal A}^{\dagger}$, ${\cal A}$, ${\cal R}^{\dagger}$ and ${\cal R}$. We should mention that the original FRpnQRPA is lacking this feature. Due to the fact that in the previous publication  the effects generated by non-commutativity of the component operators
 of the FRpnQRPA phonon operators were thrown out,
the results of the present paper constitute a step forward for the renormalization formalisms.  It is concluded that the ISR is underestimated by the general scheme, but has the advantage that keeps the standard BCS unmodified.
One hopes, however,
that adding a boson expansion (BE) on the top of the FRpnQRPA formalism, the ISR will be further improved. This expectation is supported by the results of a previous work \cite{RaSiFa} where a BE was performed in terms of the partially renormalized phonon operators (where the scattering terms are ignored). The result is determined by the fact that while the renormalized pnQRPA underestimates the ISR, the BE has an opposite effect.
 
\section{Appendix A}
Here we list the exact expressions for the commutation relations of the operators
${\cal A}^{\dag}_{1\mu}(pn), {\cal A}_{1\mu}(pn), {\bf {\bf A}}^{\dag}_{1\mu}(pn),
{\bf {\bf A}}_{1\mu}(pn)$:
\begin{eqnarray}
\left[{\cal A}_{1\mu}(pn),{\bf {\bf A}}^{\dag}_{a\mu '}(p'n')\right]&=&
\sum_{f}{\hat 1}{\hat f}C^{1\;\;f\;\;1}_{\mu\;\;m_f\;\mu '}
(-)^fW(1j_pfj_{n'};j_n1)\left[c^{\dag}_nc^{\dag}_{n'}\right]_{fm_f}\delta_{p,p'}
,\nonumber\\
\left[{\cal A}^{\dag}_{1\mu}(pn),{\bf {\bf A}}^{\dag}_{1,-\mu '}(p'n')
(-)^{1-\mu '}\right]&=&
\sum_{f}3(-)^fC^{1\;\;1\;\;f}_{\mu\;\;-\mu'\;-m_f}
W(j_nj_p1f;1j_{p'})\left[c^{\dag}_pc^{\dag}_{p'}\right]_{f,-m_f}\delta_{n,n'},
\nonumber\\
\left[{\cal A}_{1\mu}(pn),{\cal A}^{\dag}_{1\mu '}(p'n')\right]&=&
\sum_{f}{\hat 1}{\hat f}C^{1\;\;f\;\;1}_{\mu\;\;m_f\;\mu '}
\left[(-)^{f+1}W(1j_pfj_{n'};j_n 1)\left[c^{\dag}_nc_{\widetilde{n'}}\right]_{f m_f}\delta_{p,p'}
\right.\nonumber\\
&+&\left.W(1j_nfj_{p'};j_p1)\left[c^{\dag}_{p'}c_{\widetilde{p}}\right]_{fm_f}
\delta_{n,n'}\right],
\nonumber\\
\left[{\bf {\bf  A}}_{1\mu}(pn),{\bf{\bf A}}^{\dag}_{1\mu '}(p'n')\right]&=&
\sum_{f}{\hat 1}{\hat f}C^{1\;\;f\;\;1}_{\mu\;\;m_f\;\mu '}
\left[(-)^{j_n+j_{n'}+1}W(1j_pfj_{n'};j_n1)\left[c^{\dag}_{n'}c_{\widetilde{n}}\right]_{fm_f}\delta_{p,p'}
\right.\nonumber\\
&-&\left.W(1j_nfj_{p'};j_p1)\left[c^{\dag}_{p'}c_{\widetilde{p}}\right]_{fm_f}\delta_{n,n'}\right].
\end{eqnarray}

\section{Appendix B}
Here we give the explicit expressions for the coefficients $T_{ik}$ involved in Eq.\ref{EQMOT}:
\begin{eqnarray}
T_{11}(p_1n_1;pn)&=&\left[E_{p_1}\left(U^2_{p_1}-V^2_{p_1}-2U_{p_1}V_{p_1}
\frac{b_1c_1-a_1z_1}{a_1^2-b_1^2}\right)+
E_{n_1}\left(V^2_{n_1}-U^2_{n_1}-2U_{n_1}V_{n_1}
\frac{b_1z_1-a_1c_1}{a_1^2-b_1^2}\right)\right]
\nonumber\\
&\times&\delta_{p,p_1}\delta_{n,n_1}
+2\sigma^{(1)}_{p_1n_1;pn}\left[\chi-\chi_1\sigma^{(1)}_{p_1n_1;pn}
\frac{(b_1z_1-a_1c_1)(bz-ac)+(b_1c_1-a_1z_1)(bc-az)}{(a^2-b^2)(a_1^2-b_1^2)}\right],\nonumber\\
T_{12}(p_1n_1;pn)&=&\left[-2E_{p_1}U_{p_1}V_{p_1}\frac{b_1z_1-a_1c_1}{a_1^2-b_1^2}+2E_{n_1}U_{n_1}V_{n_1}\frac{b_1c_1-a_1z_1}{a_1^2-b_1^2}\right]\delta_{p,p_1}\delta_{n,n_1}\nonumber\\
&-&2\chi_1\sigma^{(1)}_{p_1n_1;pn}\frac{(b_1z_1-a_1c_1)(bc-az)+(b_1c_1-a_1z_1)(bz-ac)}{(a^2-b^2)(a_1^2-b_1^2)},\nonumber\\
T_{13}(p_1n_1;pn)&=&-2\chi_1\sigma^{(13)}_{p_1n_1;pn}\frac{(b_1z_1-a_1c_1)a-(b_1c_1-a_1z_1)b}{(a^2-b^2)(a_1^2-b_1^2)},\nonumber\\
T_{14}(p_1n_1;pn)&=&-2\chi_1\sigma^{(13)}_{p_1n_1;pn}\frac{-(b_1z_1-a_1c_1)b+
(b_1c_1-a_1z_1)a}{(a^2-b^2)(a_1^2-b_1^2)},
\nonumber\\
T_{21}(p_1n_1,pn)&=&-T_{12}^*(p_1n_1,pn),
\nonumber\\
T_{22}(p_1n_1,pn)&=&-T_{11}^*(p_1n_1,pn),
\nonumber\\
T_{23}(p_1n_1,pn)&=&-T_{14}^*(p_1n_1,pn),
\nonumber\\
T_{24}(p_1n_1,pn)&=&-T_{13}^*(p_1n_1,pn),
\end{eqnarray}
\begin{eqnarray}
&&T_{31}(p_1n_1,pn)=\frac{|D_3(p_1n_1)|^{-1/2}D^{1/2}_1(pn)}{a_1^2-b_1^2}\left[E_{p_1}\left[\left((U_{p_1}^2-V_{p_1}^2)c_1+2U_{p_1}V_{p_1}b_1\right)
(a_1^2-b_1^2)\right.\right.\nonumber\\
&+&\left.\left.(b_1z_1-a_1c_1)\left((U_{p_1}^2-V_{p_1}^2)a_1-2U_{p_1}V_{p_1}z_1\right)+(b_ac_1-a_1z_1)\left((U_{p_1}^2-V_{p_1}^2)b_1-2U_{p_1}V_{p_1}c_1\right)\right]\right.+
\nonumber\\
&+&\left.E_{n_1}\left[\left((V_{n_1}^2-U_{n_1}^2)c_1+2U_{n_1}V_{n_1}a_1\right)
(a_1^2-b_1^2)
+(b_1z_1-a_1c_1)\left((U_{n_1}^2-V_{n_1}^2)a_1+2U_{n_1}V_{n_1}c_1\right)
\right.\right.
\nonumber\\
&+&\left.\left.(b_1c_1-a_1z_1)\left((U_{n_1}^2-V_{n_1}^2)b_1+2U_{n_1}V_{n_1}z_1\right)\right]\right]\delta_{p_1,p}\delta_{n_1,n}\nonumber\\
&-&2\chi_1\sigma^{(31)}(p_1n_1;pn)\frac{(bz-ac)a_1-(bc-az)b_1}{(a^2-b^2)(a_1^2-b_1^2)},
\nonumber\\
&&T_{32}(p_1n_1,pn)=\frac{|D_3(p_1n_1)|^{-1/2}D^{1/2}_1(pn)}{a_1^2-b_1^2}
\left[E_{p_1}\left[\left((U_{p_1}^2-V_{p_1}^2)z_1+2U_{p_1}V_{p_1}a_1\right)
(a_1^2-b_1^2)\right.\right.  \nonumber\\
&+&\left.\left.(b_1z_1-a_1c_1)\left((U_{p_1}^2-V_{p_1}^2)b_1-2U_{p_1}V_{p_1}c_1\right)+(b_1c_1-a_1z_1)\left((U_{p_1}^2-V_{p_1}^2)a_1-2U_{p_1}V_{p_1}z_1\right)\right]\right.+\nonumber\\
&+&\left.E_{n_1}\left[\left((V_{n_1}^2-U_{n_1}^2)z_1+2U_{n_1}V_{n_1}b_1\right)
(a_1^2-b_1^2)+(b_1z_1-a_1c_1)\left((U_{n_1}^2-V_{n_1}^2)b_1+2U_{n_1}V_{n_1}z_1\right)\right.\right.
\nonumber\\
&+&\left.\left.(b_1c_1-a_1z_1)\left((U_{n_1}^2-V_{n_1}^2)a_1+2U_{n_1}V_{n_1}c_1\right)\right]\right]\delta_{p_1,p}\delta_{n_1,n}\nonumber\\
&-&2\chi_1\sigma^{(31)}(p_1n_1;pn)\frac{(bc-az)a_1-(bz-ac)b_1}{(a^2-b^2)(a_1^2-b_1^2)},
\nonumber\\
&&T_{33}(p_1n_1,pn)=\frac{\delta_{p_1,p}\delta_{n_1,n}}{a_1^2-b_1^2}
\left[E_{p_1}\left((U_{p_1}^2-V_{p_1}^2)(a_1^2-b_1^2)-2U_{p_1}V_{p_1}
(a_1z_1-b_1c_1)\right)\right.
\nonumber\\
&+&\left.E_{n_1}\left((U_{n_1}^2-V_{n_1}^2)(a_1^2-b_1^2)+2U_{n_1}V_{n_1}
(a_1c_1-b_1z_1)\right)\right]\nonumber\\
&-&2\chi_1\sigma^{(3)}(p_1n_1;pn)\frac{aa_1+bb_1}{(a^2-b^2)(a_1^2-b_1^2)},
\nonumber\\
&&T_{34}(p_1n_1,pn)=\frac{\delta_{p_1,p}\delta_{n_1,n}}{a_1^2-b_1^2}
\left[2E_{p_1}U_{p_1}V_{p_1}(b_1z_1-a_1c_1)+
2E_{n_1}U_{n_1}V_{n_1}(a_1z_1-b_1c_1)\right]\nonumber\\
&+&2\chi_1\sigma^{(3)}(p_1n_1;pn)\frac{ba_1+ab_1}{(a^2-b^2)(a_1^2-b_1^2)},
\nonumber\\
&&T_{41}(p_1n_1,pn)=-T_{32}^*(p_1n_1,pn),
\nonumber\\
&&T_{42}(p_1n_1,pn)=-T_{31}^*(p_1n_1,pn),
\nonumber\\
&&T_{43}(p_1n_1,pn)=-T_{34}^*(p_1n_1,pn),
\nonumber\\
&&T_{44}(p_1n_1,pn)=-T_{33}^*(p_1n_1,pn).
\end{eqnarray}
The expressions for $a,b,c,z$ have been defined before by Eq.(4.3).
The coefficients $a_1,b_1,c_1,z_1$ are obtained from 
Eq.(4.3) by replacing the indices $p$ and $n$ by $p_1$ and $n_1$, respectively. 
The index $^*$ stands for the complex conjugation.
Several notations have been used:
\begin{eqnarray}
\sigma^{(1)}(p_1n_1,pn)&=&D_1^{1/2}(p_1n_1)\sigma_{p_1n_1,pn}D_1^{1/2}(pn),\nonumber\\
\sigma^{(3)}(p_1n_1,pn)&=&\epsilon_{p_1n_1}|D_3(p_1n_1)|^{1/2}\sigma_{p_1n_1,pn}|D_3(pn)|^{1/2},\nonumber\\
\sigma^{(13)}(p_1n_1,pn)&=&D_1^{1/2}(p_1n_1)\sigma_{p_1n_1,pn}|D_3(pn)|^{1/2},\nonumber\\
\sigma^{(31)}(p_1n_1,pn)&=&\epsilon_{p_1n_1}|D_3(p_1n_1)|^{1/2}\sigma_{p_1n_1,pn}
D_1^{1/2}(pn),\nonumber\\
\epsilon_{p_1n_1}&=&\frac{D_3(p_1n_1)}{|D_3(p_1n_1)|}.
\end{eqnarray}
\section{Appendix C}
Here we give the explicit expressions for the bar amplitudes involved in the equations for the
quasiparticle number operators averages (\ref{NpNn1}). The results are:
\begin{eqnarray}
\bar{X}_k(pn)&=&\frac{1}{\sqrt{D_1(pn)}}\left(U_pV_nX_k(pn)-U_nV_pY_k(pn)\right)+
\frac{1}{\sqrt{|D_3(pn)|}}\nonumber\\
&&\left[Z_k(pn)\left(aU_pU_n-bV_pV_n+cU_pV_n+zU_nV_p\right)\right.\nonumber\\
&-&\left. W_k(pn)\left(-aV_pV_n+bU_pU_n+cU_nV_p+zU_pV_n\right)\right],\nonumber\\
-\bar{Y}_k(pn)&=&\frac{1}{\sqrt{D_1(pn)}}\left(U_nV_pX_k(pn)-U_pV_nY_k(pn)
\right)+
\frac{1}{\sqrt{|D_3(pn)|}}\nonumber\\
&&\left[Z_k(pn)\left(-aV_pV_n+bU_pU_n+cU_nV_p+zU_pV_n\right)\right.\nonumber\\
&-&\left. W_k(pn)\left(aU_pU_n-bV_pV_n+cU_pV_n+zU_nV_p\right)\right],\nonumber\\
\bar{Z}_k(pn)&=&\frac{1}{\sqrt{D_1(pn)}}\left(U_pU_nX_k(pn)+V_nV_pY_k(pn)\right)+
\frac{1}{\sqrt{|D_3(pn)|}}\nonumber\\
&&\left[Z_k(pn)\left(-aU_pV_n-bV_pU_n+cU_pU_n-zV_nV_p\right)\right.\nonumber\\
&-&\left. W_k(pn)\left(-aV_pU_n-bU_pV_n-cV_nV_p+zU_pU_n\right)\right],\nonumber\\
-\bar{W}_k(pn)&=&\frac{1}{\sqrt{D_1(pn)}}\left(-V_nV_pX_k(pn)-U_pU_nY_k(pn)\right)+
\frac{1}{\sqrt{|D_3(pn)|}}\nonumber\\
&&\left[ Z_k(pn)\left(-aV_pU_n-bU_pV_n-cV_nV_p+zU_pU_n\right)\right.\nonumber\\
&-&\left. W_k(pn)\left(-aU_pV_n-bV_pU_n+cU_pU_n-zV_nV_p\right)\right].
\end{eqnarray}


\begin{references}
\bibitem{GreiMar}W.Greiner and J. Maruhn, Nuclear Models (Springer Verlag, Berlin, Heidelberg, New York).
\bibitem{Ring}P.Ring and P.Schuck, The Nuclear Many-body Problem (Springer, New York, 1980).
\bibitem{Machia}A.O. Machiavelli, Z. Harris and P. Fallon, International, Nuclear Physics Conference, INPC 2001, eds. Norman {\it at al.} (AIP 2002),page 783.
\bibitem{Pov}C.A. Ur {\it et al.} Phys. Rev. {\bf C58} (1998) 3163; C.E. Svensson {\it et al.} Phys. Rev. {\bf C 58} (1998) R2621.
\bibitem{Su}J. Suhonen and O. Civitarese, Phys. Rep. {\bf 300} (1988) 123.
\bibitem{Fa}A. Faessler, Prog. Part. Nucl. Phys. {\bf 21} (1988) 183.
\bibitem{Cha} D. Cha, Phys. Rev. {\bf C27}(1983) 2269.
\bibitem{Rad1}A.A.Raduta,A.Faessler, S. Stoica and W. Kaminski, Phys. Lett. {\bf B254} (1991) 7.
\bibitem{Rad2}A.A.Raduta, A. Faessler and S. Stoica, Nucl. Phys. {\bf A 534} (1991) 149.
\bibitem{Su1}J.Toivanen, J. Suhonen, Phys. Rev. Lett. {\bf 75} (1995) 410.
\bibitem{Rad3}A.A. Raduta, C. M. Raduta, A. Faessler and W. Kaminski, Nucl. Phys. {\bf A634}
(1998) 497.
\bibitem{Bla}J.P.Blaizot and E.R.Marshalek, Nucl. Phys. {\bf A309} (1978) 493.
\bibitem{Rad4} A.A. Raduta, C. M. Raduta, B. Codirla, Nucl. Phys. {\bf 678} (2000) 382.
\bibitem{Ike}K.Ikeda, Prog. Theor. Phys. {\bf 31} (1964) 434.
\bibitem{Ro}V. Rodin and A.Faessler, Phys. Rev. {\bf C66} (2002) 051303(R).
\bibitem{Schu}J. Dukelsky, G. Roepke and P. Shuck, Nucl. Phys. {\bf A628} (1998) 17.
\bibitem{Rose}M.E.Rose, Elementary Theory of Angular Momentum (Wiley, New York, 1957).
\bibitem{Jol}R.V.Jolos and W. Rybarska-Nawrocka, Z. Physik {\bf A 296} (1980) 73.
\bibitem{Kri}F. Krmpotic et al. Nucl. Phys. {\bf A637} (1998) 295.
\bibitem{Boby}A. Bobyk, W. A. Kaminsky, P. Zareba, The Eur. Phys. Jour. {\bf A5} (1999) 385.
\bibitem{RaSiFa} A.A.Raduta, F. Simkovich and A. Faessler, Jour. Phys. {\bf G26} (2000) 793.
\end{references}
\end{document}